\documentclass[aps,prl,reprint,superscriptaddress]{revtex4-1}

\usepackage{graphicx}
\usepackage{multibib}
\newcites{S}{Supplementary}

\begin{document}


\title{Fast spin-orbit qubit in an indium antimonide nanowire}



\author{J.W.G. \surname{van den Berg}}
\affiliation{Kavli Institute of Nanoscience, Delft University of Technology, 2600 GA Delft, The Netherlands}
\author{S. \surname{Nadj-Perge}}
\affiliation{Kavli Institute of Nanoscience, Delft University of Technology, 2600 GA Delft, The Netherlands}
\author{V.S. Pribiag}
\affiliation{Kavli Institute of Nanoscience, Delft University of Technology, 2600 GA Delft, The Netherlands}
\author{S.R. Plissard}
\affiliation{Department of Applied Physics, Eindhoven University of Technology, 5600 MB Eindhoven, The Netherlands}
\author{E.P.A.M. Bakkers}
\affiliation{Kavli Institute of Nanoscience, Delft University of Technology, 2600 GA Delft, The Netherlands}
\affiliation{Department of Applied Physics, Eindhoven University of Technology, 5600 MB Eindhoven, The Netherlands}
\author{S.M. Frolov}
\affiliation{Kavli Institute of Nanoscience, Delft University of Technology, 2600 GA Delft, The Netherlands}
\author{L.P. Kouwenhoven}
\affiliation{Kavli Institute of Nanoscience, Delft University of Technology, 2600 GA Delft, The Netherlands}


\date{\today}

\begin{abstract}
Due to the strong spin-orbit interaction in indium antimonide, orbital motion and spin are no longer separated. This enables fast manipulation of qubit states by means of microwave electric fields.
We report Rabi oscillation frequencies exceeding 100~MHz for spin-orbit qubits in InSb nanowires. Individual qubits can be selectively addressed due to intrinsic differences in their $g$-factors. 
Based on Ramsey fringe measurements, we extract a coherence time $T_2^*$ = $8\pm1$~ns at a driving frequency of 18.65~GHz.
Applying a Hahn echo sequence extends this coherence time to 35~ns.
\end{abstract}

\pacs{}

\maketitle


The spin of a single electron forms a two-level system, which makes it a natural choice for creating a quantum bit (qubit) \cite{Loss1998}. Quantum information processing based on such qubits has developed into a mature and diverse field \cite{Hanson2007}. Previous work has demonstrated important milestones, including single-shot detection of spin state, coherent control of a single spin and coherent coupling between two spins \cite{Hanson2007, Petta2005a, Simmons2011, Nadj-Perge2010a, Maune2012}. Of great importance for future development of spin-based quantum computation is combining efficient single-qubit control and two-qubit operations in the same system \cite{Shulman2012} and developing ways to integrate spin qubits with other quantum computing architectures. To pursue these goals, several promising material platforms are being explored. Among these are narrow band-gap semiconductor nanowires, such as indium arsenide and indium antimonide. This class of materials has recently gained considerable attention, due to their strong spin-orbit coupling, which enables efficient all-electrical spin control \cite{Bjork2005, Pfund2007a, Nilsson2009, Nadj-Perge2010a, Nadj-Perge2010} and could provide a means of coupling qubits to quantum systems based on superconducting cavities \cite{Petersson2012}.

In this paper we demonstrate an electrically controlled spin-orbit qubit in an indium antimonide nanowire. 
We observe Rabi oscillations with frequencies up to 104 MHz, the fastest reported to date for an electrically controlled single-spin qubit in a quantum dot. Furtermore, we show that the individual qubits in the two dots can be addressed with high selectivity, owing to a large g-factor difference between two dots. 
We achieve universal qubit control and study qubit coherence by means of Ramsey type measurements. We find that the inhomogeneous dephasing time $T_2^*$ can be extended to $\sim 35$~ns by using a Hahn echo.

To realize our spin-orbit qubit, a double quantum dot is defined inside the nanowire by means of local electrostatic gating. The qubits' basis states are spin-orbit doublets (denoted by $\Uparrow$ and $\Downarrow$), which---analogous to conventional spin qubits---are split by the Zeeman energy in a magnetic field. Transitions between these states can be induced by applying microwave frequency electric fields.

\begin{figure}[hbt]
\includegraphics{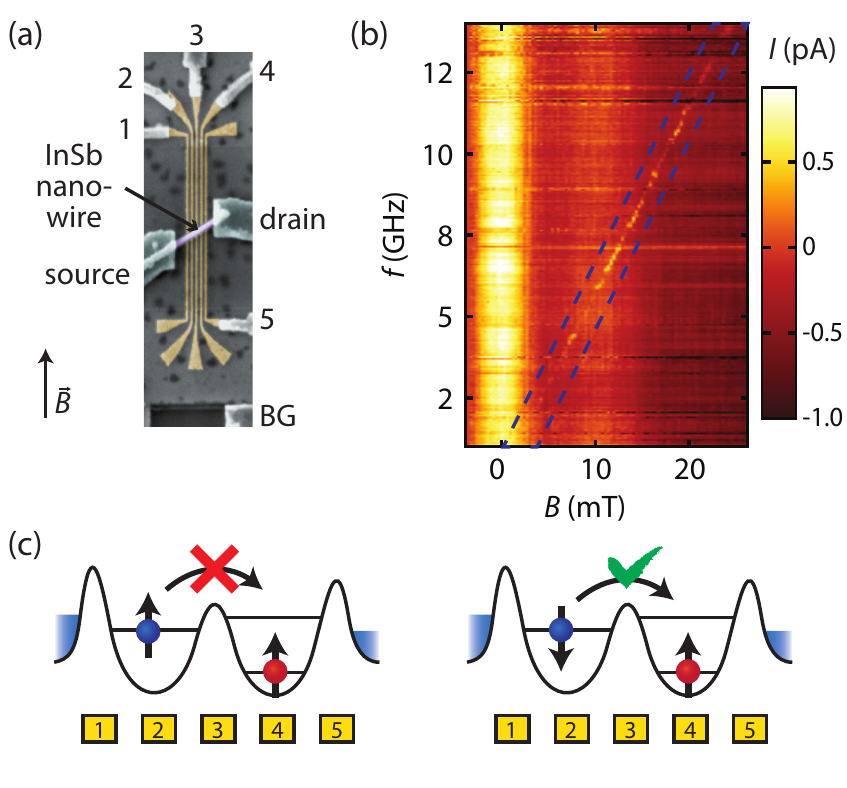}%
\caption{\label{fig1}(color online) (a) Electron microscope image of the device consisting of an InSb nanowire contacted by source and drain electrodes, lying across a set of fine gates (numbered, 60 nm pitch) as well as a larger bottom gate (labeled BG).
(b) Current through the device when applying microwaves with the double dot in spin blockade configuration (a vertical linecut near 0~mT has been subtracted to suppress resonances at constant frequency). When the microwave frequency matches the Larmor frequency (resonance highlighted by dashed blue box), blockade is lifted and current increases. 
(c) Schematic illustration of Pauli spin blockade on which read-out depends. Only anti-parallel states (right) can occupy the same dot, allowing current through the device. A parallel configuration (left) leads to a suppression of the current.
}
\end{figure}

An image of our device obtained by scanning electron microscopy is presented in figure \ref{fig1}(a). It consists of an indium antimonide nanowire ($\sim 1.5 \mu$m long, ~100 nm thick) contacted by Ti/Al source and drain electrodes. Below the nanowire, separated by a layer of Si$_3$N$_4$ dielectric, is a set of 5 narrow gates (60 nm pitch) used to induce a double quantum dot potential in the nanowire and control the number of electrons in these dots. Underneath another layer of Si$_3$N$_4$ is a large metallic gate (BG) by which the conductance of the entire wire can be tuned. Measurements are performed in a $^3$He system with a base temperature of 260~mK.

In order to read out and initialize our qubits we take advantage of Pauli spin blockade \cite{Hanson2007, Pfund2007}. When a bias voltage is applied between the source and drain of the double dot, sequential transport through the dots is possible in a triangular region in gate space. However, spin conservation introduces additional constraints on interdot tunneling. Although a transition may be energetically allowed, it may be prohibited by spin selection rules. For example, a triplet state with one electron in each dot such as ($\Uparrow,\Uparrow$) cannot transition to a (0,2) singlet state, unless one of the spin-orbit states is rotated. The current through the dots thus becomes supressed as eventually such a parallel triplet state is loaded. This principle is illustrated in figure \ref{fig1}(c). Note that the anti-parallel $T_0$ state is in practice not blocked as hyperfine interaction makes it quickly decay into a singlet state \cite{Hanson2007}.
Rotation of a spin-orbit state from $\Uparrow$ to $\Downarrow$, or vice-versa, can be accomplished by electric-dipole spin resonance (EDSR). Application of an alternating electric field, resonant with the Larmor precession frequency, drives transitions between the spin-orbit states \cite{Rashba2003a, Golovach2006, Laird2007, Pioro-Ladriere2008}, lifting the spin blockade.
To detect the EDSR, we measure the spin blockade leakage current as a function of applied magnetic field, $B$, and microwave frequency, $f$. (Fig. \ref{fig1}(b)). From the slope of the resonance we extract a Land\'e $g$-factor of 41 for our quantum dot. In addition to the resonance we observe a finite current around $B=0$ \footnote{Due to a small offset in the magnet power supply output, the magnetic field scale has been offset to have 0 coincide with the hyperfine peak.}, resulting from mixing of the (1,1) triplet states to the singlet by the nuclear field \cite{Koppens2005a,Nadj-Perge2012}.

\begin{figure}
\includegraphics{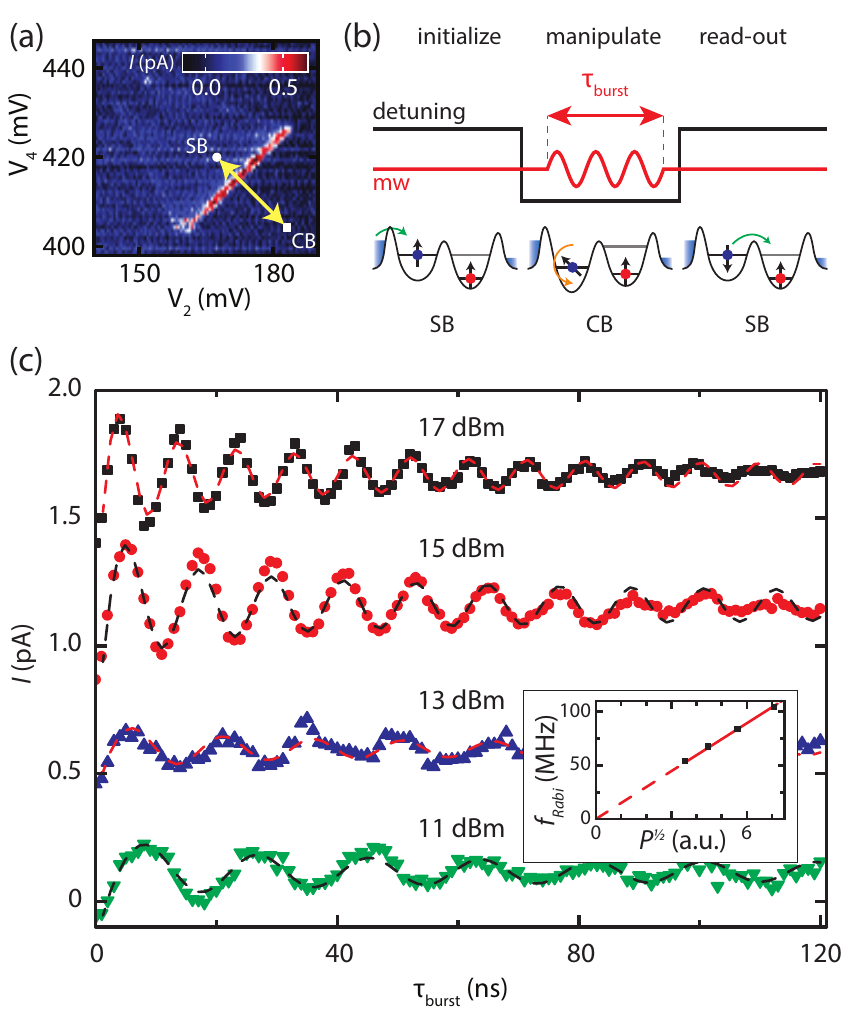}%
\caption{\label{fig2}(color online) (a) Bias triangle in which spin blockade was observed for a negative bias of -5~mV. This is the ($2m+1$,$2n+1$)$\rightarrow$($2m$,$2n+2$) transition (transition A, see \cite{supmat}).
(b) Sequence used for measuring Rabi oscillations. Pulses are applied to gates 2 and 4 to move the double dot along the detuning axis between Coulomb blockade (CB) and spin blockade (SB) configurations. In CB a microwave burst is applied via gate BG to rotate the spin. 
(c) Rabi oscillation obtained at a driving frequency of 18.65 GHz and source power of 11 (bottom) to 17 (top) dBm. Dashed lines are fits to $a \cos(f_R \tau_{burst}+\phi)\tau_{burst}^{-d}+b$, giving Rabi frequencies $f_R$ of $54 \pm 1$; $67 \pm 1$; $84 \pm1$ and $104 \pm 1$ MHz. Linear slopes, attributed to photon assisted tunneling, of 0.5, 0.6, 0.4, and 0.6 fA/ns (top to bottom) were subtracted. $d = 0.5$ for the bottom trace and 0.4 for the others. Curves are offset by 0.5 pA for clarity.
Inset: Rabi frequencies set out against driving amplitudes, including a linear fit through 0.
}
\end{figure}

To demonstrate coherent control over the qubit we apply microwave bursts of variable length. First, the qubit is initialized into a spin blocked charge configuration. This is accomplished by idling inside the bias triangle (Fig. \ref{fig2}(a)). In order to prevent the electron from tunneling out of the dot during its subsequent manipulation, the double dot is maintained in Coulomb blockade in the same charge configuration. While in the Coulomb blockade regime, a microwave burst is applied. The double dot is then again quickly brought back to the spin blockade configuration by pulsing the plunger gates. By applying such microwave bursts (schematically depicted in figure \ref{fig2}(b)), we perform a Rabi measurement. If the manipulation has flipped the electron spin-orbit state, the blockade is lifted and an electron can move from the first to the second dot and exit again through the outgoing lead. By continuously repeating the pulse sequence and measuring the (DC) current through the double dot, we measure the Rabi oscillations associated with the rotation of the spin-orbit state (Fig. \ref{fig2}(c)).

We have achieved Rabi oscillations with more than five oscillation periods at a maximum frequency of $104$~MHz. Even higher Rabi frequencies could be observed for larger microwave driving power \cite{supmat}, but this resulted in a quick damping of the oscillations, likely due to photon assisted tunneling (PAT) \cite{Koppens2006a,Nowack2007a}. The high frequency allows for fast operation of the qubit, which is important for quantum information processing.
Also important in this respect is the qubit manipulation fidelity. We can estimate this fidelity from the strength of the spin-orbit field relative to the fluctuations of the nuclear spin bath (believed to be the main source of dephasing) \cite{Koppens2006a}. From the Rabi frequency of 104~MHz we obtain a spin-orbit field $B_{SO} = 0.36$~mT driving the rotations. By taking the width of the EDSR peak as a measure for the nuclear magnetic field we can estimate a value of $B_N = 0.16 \pm 0.02$~mT for this field. With these values of $B_N$ and $B_{SO}$ we estimate our qubit manipultion fidelity to be $81 \pm 6$~\% \cite{supmat}.

\begin{figure}
\includegraphics[width=\columnwidth]{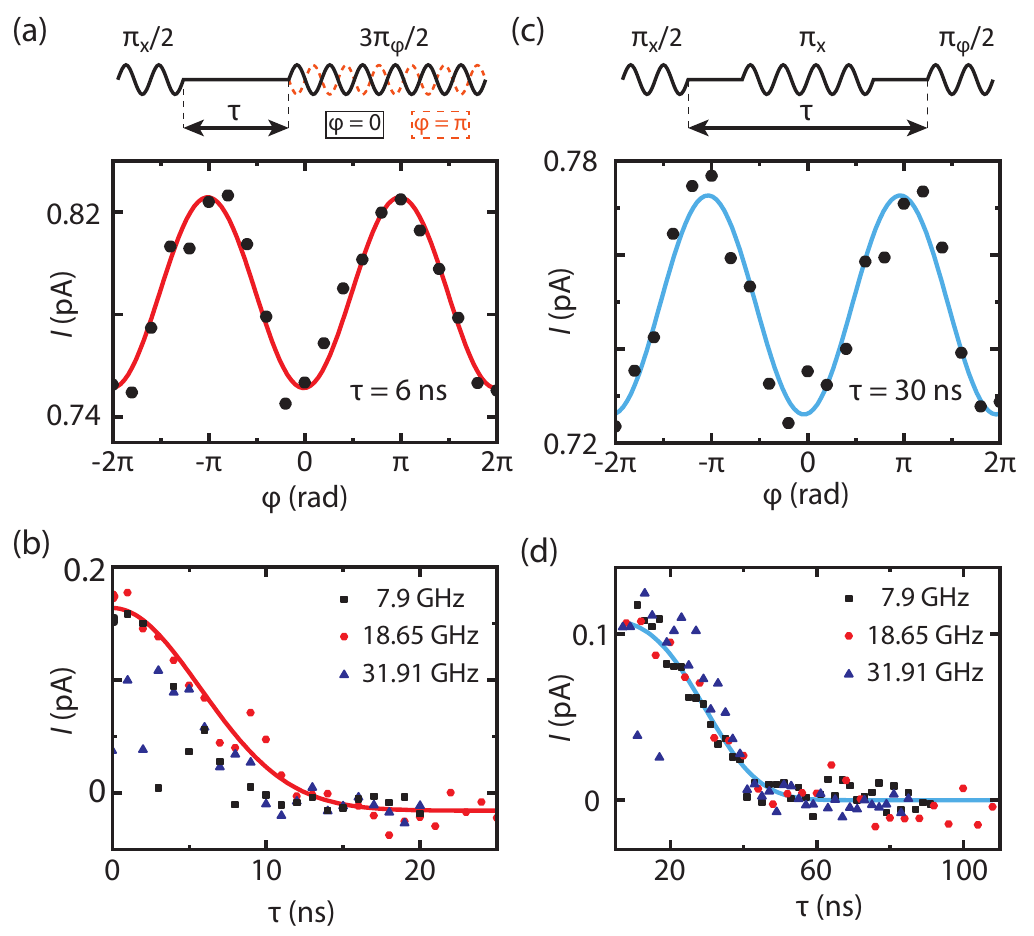}%
\caption{\label{fig3}(color online) (a) Ramsey experiment; an initial $\pi/2$-pulse rotates the spin to the xy-plane. After some delay a $3\pi/2$ pulse is applied, restoring spin blockade or (partially) lifting it, depending on the phase of the pulse.
(b) Decay of the Ramsey fringe contrast with increasing delay time $\tau$ for different driving frequencies. Solid line is a fit to $\exp(-(\tau/T_2^*)^2)$ at a driving frequency of 18.65~GHz, giving $T_2^* = 8 \pm 1$ ns.
(c) A Hahn echo sequence (top), extends the decay of the fringe contrast, to 30~ns in this case.
(d) Decay of the fringe contrast in the Hahn echo sequence for different microwave frequencies. Solid line is a fit to $\exp(-(\tau/T_{echo})^3)$ for driving frequency 18.65 GHz, yielding $T_{echo} = 34 \pm 2$ ns.}
\end{figure}

The Rabi experiment demonstrates rotation of the qubit around a single axis. However, in order to be able to prepare the qubit in any arbitrary superposition, it is necessary to achieve rotations around two independent axes. We demonstrate such universal control by means of a Ramsey experiment, where the axis of qubit rotation is determined by varying the phase of the applied microwave bursts, as illustrated at the top of figure \ref{fig3}(a).
As in the Rabi experiment, the microwave bursts are applied while the dots are kept in Coulomb blockade, to maintain a well defined charge state and prevent the electrons from tunneling out during manipulation. In the Ramsey sequence an initial microwave burst rotates the state by $\pi/2$ to the xy-plane of the Bloch sphere. We take this rotation axis to be the x-axis. A second burst is then applied after some delay $\tau$, making a $3\pi/2$ rotation. By varying the phase of this pulse with respect to the initial $\pi/2$ pulse, we can control the axis of the second rotation (see figure \ref{fig3}(a)). For example, if the two bursts are applied with the same phase, in total a $2 \pi$ rotation will have been made. This restores a spin blockade configuration, thus leading to a suppression of the current. A second burst with a phase $\pi$, however, would rotate the qubit in the opposite direction, ending up along the $\left| -z \right\rangle$-direction on the Bloch sphere. For this case spin blockade is thus fully lifted and current increases to a maximum.

When the delay time between the first and final pulse in the Ramsey sequence is increased, the qubit starts to dephase.
The loss of phase coherence leads to decay of the Ramsey fringe contrast, as shown in figure \ref{fig3}(b). By fitting the experimental data to $\exp(-(\tau/T_2^*)^2)$ we extract a dephasing time of $T_2^* = 8 \pm 1$ ns, obtained at a driving frequency of 18.65 GHz. Other driving frequencies of 7.9~GHz and 31.91~GHz resulted in similar $T_2^*$ values of $6\pm1$ and $9\pm1$ respectively. 
To extend the coherence of the qubit, we employ a Hahn echo technique \cite{Herzog1956, Koppens2008}: halfway between two $\pi/2$ pulses an extra pulse is applied to flip the state over an angle $\pi$. Doing so partially refocuses the dephasing caused by the nuclear magnetic field, which varies slowly compared to the electron spin dynamics \cite{Reilly2008a}. From figure \ref{fig3}(c), where the total delay has been extended to $\tau = 30$~ns, it is clear that this technique can maintain contrast of the Ramsey fringes for considerably longer times. An increase in the coherence time to $T_{echo} = 35 \pm 1$~ns is obtained from the decay of the contrast (figure \ref{fig3}(d)) for a driving frequency of 18.65 GHz. Similar values of $34 \pm 2$ and $32\pm1$~ns were obtained at driving frequencies 7.9 and 31.91~GHz respectively.

The relatively low coherence times obtained here are in line with previous results obtained for InAs nanowires \cite{Nadj-Perge2010}. The timescales show no significant dependence on driving frequency within the accessible range of 8--32~GHz. These results suggest the existence of a fast spin bath interacting with the electron, likely originating from the large nuclear spins of indium antimonide (5/2 and 7/2 for $^{121}$Sb and $^{123}$Sb respectively, and 9/2 for In). It must be noted though that other sources of dephasing, such as nearby paramagnetic impurities or charge noise, cannot be completely ruled out.

\begin{figure}
\includegraphics{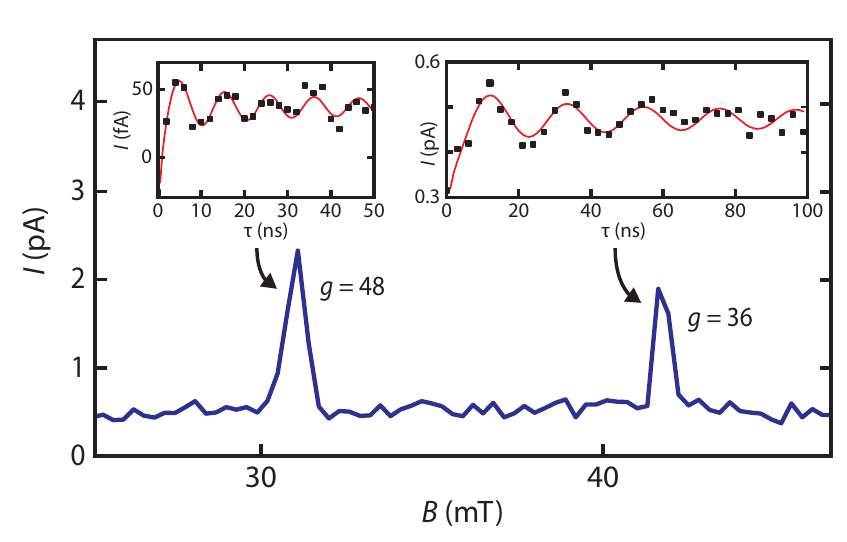}%
\caption{\label{fig4}(color online) Main panel: Two well separated EDSR peaks for the spin-orbit qubit in each of the two dots. The microwave driving frequency is 20.9~GHz.
Insets: Rabi oscillations for the corresponding EDSR peaks. Linear slopes (attributed to PAT) of 0.6 and 0.9~fA/ns respectively are subtracted to flatten the average. The Rabi data on the left was obtained at 31.2~mT $B$-field and 20.9~GHz driving frequency. From the fit (as in Fig. \ref{fig2}) a Rabi frequency of $96 \pm 2$~MHz is obtained. On the right the field was 41.2~mT and driving frequency 21~GHz. 
The Rabi frequency obtained from the fit is $47 \pm 3$ MHz.
}
\end{figure}

Thus far we have only presented coherent rotations of one of the qubits. It is possible to individually address each qubit in the two dots if the EDSR peak splits into two separate resonances. This splitting can arise due to a $g$-factor difference between the two dots, as in our present device, but can also be engineered, e.g. through incorporation of a micromagnet into a device \cite{Pioro-Ladriere2008, Tokura2006}. 
The main panel of figure \ref{fig4} displays the two EDSR peaks corresponding to the two dots. From the magnetic field dependence of the resonances, we determine $g$-factors of 48 and 36 for the two dots respectively \cite{supmat}.
This difference in $g$-factors for the two dots can  be explained by a difference in confinement \cite{Pryor2006, Nadj-Perge2010, Nadj-Perge2012, Jung2011a}. Note that these measurements were obtained at the charge transition ($2m+1$,$2n+3$)$\rightarrow$($2m+2$,$2n+2$) (transition B, \cite{supmat}) in opposite bias than the previously presented data. No second resonance was resolved for charge transition A, perhaps due to strongly similar $g$-factors, or decreased coupling of the microwaves to the second dot.

Utilizing the large difference in $g$-factors between the two dots we have achieved coherent control of both qubits. Here, we probe the qubits using similar microwave frequencies, but different magnetic fields. The insets in figure \ref{fig4} show the Rabi oscillations obtained for each of the qubits. The frequency of the Rabi oscillations (see insets of figure \ref{fig4}) for the qubit corresponding to $g$-factor 48, was 96~MHz. For the other dot, with a corresponding $g$-factor of 36, a lower Rabi frequency of 47~MHz was achieved. This slower Rabi oscillation is consistent with a weaker coupling of the microwave electric field to this dot. \footnote{The Rabi for the $g = 36$ dot was achieved at different gate voltages than for the other dot, to improve its visibility, though care was taken to measure the intended resonance.}

In summary, we have demonstrated the realization of a spin-orbit qubit in an indium antimonide nanowire. Fast manipulation and universal control of the qubit was demonstrated by Rabi and Ramsey measurements. A significant improvement in terms of speed and fidelity was achieved compared to previously realized indium arsenide spin-orbit qubits. Dephasing times however, remained similar to InAs, suggesting that the nuclear spin bath is the main source of dephasing. The large interdot $g$-factor differences make possible the selective addressing of different qubits. Importantly, we estimate that the large interdot Zeeman energy difference would be sufficient to implement a fast ($\sim 1$~ns) two-qubit CPhase gate \cite{Burkard1999, Meunier2011}, operating well within the $T_{echo}$ coherence time. 

\begin{acknowledgments}
We would like to thank V.E. Calado for discussions. Financial support for this work was given by ERC and The Netherlands Organisation for Scientific Research NWO/FOM.
\end{acknowledgments}

\bibliography{manuscript}

\begin{thebibliography}{6}%
\makeatletter
\providecommand \@ifxundefined [1]{%
 \@ifx{#1\undefined}
}%
\providecommand \@ifnum [1]{%
 \ifnum #1\expandafter \@firstoftwo
 \else \expandafter \@secondoftwo
 \fi
}%
\providecommand \@ifx [1]{%
 \ifx #1\expandafter \@firstoftwo
 \else \expandafter \@secondoftwo
 \fi
}%
\providecommand \natexlab [1]{#1}%
\providecommand \enquote  [1]{``#1''}%
\providecommand \bibnamefont  [1]{#1}%
\providecommand \bibfnamefont [1]{#1}%
\providecommand \citenamefont [1]{#1}%
\providecommand \href@noop [0]{\@secondoftwo}%
\providecommand \href [0]{\begingroup \@sanitize@url \@href}%
\providecommand \@href[1]{\@@startlink{#1}\@@href}%
\providecommand \@@href[1]{\endgroup#1\@@endlink}%
\providecommand \@sanitize@url [0]{\catcode `\\12\catcode `\$12\catcode
  `\&12\catcode `\#12\catcode `\^12\catcode `\_12\catcode `\%12\relax}%
\providecommand \@@startlink[1]{}%
\providecommand \@@endlink[0]{}%
\providecommand \url  [0]{\begingroup\@sanitize@url \@url }%
\providecommand \@url [1]{\endgroup\@href {#1}{\urlprefix }}%
\providecommand \urlprefix  [0]{URL }%
\providecommand \Eprint [0]{\href }%
\providecommand \doibase [0]{http://dx.doi.org/}%
\providecommand \selectlanguage [0]{\@gobble}%
\providecommand \bibinfo  [0]{\@secondoftwo}%
\providecommand \bibfield  [0]{\@secondoftwo}%
\providecommand \translation [1]{[#1]}%
\providecommand \BibitemOpen [0]{}%
\providecommand \bibitemStop [0]{}%
\providecommand \bibitemNoStop [0]{.\EOS\space}%
\providecommand \EOS [0]{\spacefactor3000\relax}%
\providecommand \BibitemShut  [1]{\csname bibitem#1\endcsname}%
\let\auto@bib@innerbib\@empty
\bibitem [{\citenamefont {Jouravlev}\ and\ \citenamefont
  {Nazarov}(2006)}]{Jouravlev2006}%
  \BibitemOpen
  \bibfield  {author} {\bibinfo {author} {\bibfnamefont {O.~N.}\ \bibnamefont
  {Jouravlev}}\ and\ \bibinfo {author} {\bibfnamefont {Y.~V.}\ \bibnamefont
  {Nazarov}},\ }\href {\doibase 10.1103/PhysRevLett.96.176804} {\bibfield
  {journal} {\bibinfo  {journal} {Physical Review Letters}\ }\textbf {\bibinfo
  {volume} {96}},\ \bibinfo {pages} {176804} (\bibinfo {year}
  {2006})}\BibitemShut {NoStop}%
\bibitem [{\citenamefont {Pfund}\ \emph {et~al.}(2007)\citenamefont {Pfund},
  \citenamefont {Shorubalko}, \citenamefont {Ensslin},\ and\ \citenamefont
  {Leturcq}}]{Pfund2007S}%
  \BibitemOpen
  \bibfield  {author} {\bibinfo {author} {\bibfnamefont {A.}~\bibnamefont
  {Pfund}}, \bibinfo {author} {\bibfnamefont {I.}~\bibnamefont {Shorubalko}},
  \bibinfo {author} {\bibfnamefont {K.}~\bibnamefont {Ensslin}}, \ and\
  \bibinfo {author} {\bibfnamefont {R.}~\bibnamefont {Leturcq}},\ }\href
  {\doibase 10.1103/PhysRevLett.99.036801} {\bibfield  {journal} {\bibinfo
  {journal} {Physical Review Letters}\ }\textbf {\bibinfo {volume} {99}},\
  \bibinfo {pages} {036801} (\bibinfo {year} {2007})}\BibitemShut {NoStop}%
\bibitem [{\citenamefont {Danon}\ and\ \citenamefont
  {Nazarov}(2009)}]{Danon2009a}%
  \BibitemOpen
  \bibfield  {author} {\bibinfo {author} {\bibfnamefont {J.}~\bibnamefont
  {Danon}}\ and\ \bibinfo {author} {\bibfnamefont {Y.~V.}\ \bibnamefont
  {Nazarov}},\ }\href {\doibase 10.1103/PhysRevB.80.041301} {\bibfield
  {journal} {\bibinfo  {journal} {Physical Review B}\ }\textbf {\bibinfo
  {volume} {80}},\ \bibinfo {pages} {041301} (\bibinfo {year}
  {2009})}\BibitemShut {NoStop}%
\bibitem [{\citenamefont {Nadj-Perge}\ \emph {et~al.}(2012)\citenamefont
  {Nadj-Perge}, \citenamefont {Pribiag}, \citenamefont {van~den Berg},
  \citenamefont {Zuo}, \citenamefont {Plissard}, \citenamefont {Bakkers},
  \citenamefont {Frolov},\ and\ \citenamefont {Kouwenhoven}}]{Nadj-Perge2012S}%
  \BibitemOpen
  \bibfield  {author} {\bibinfo {author} {\bibfnamefont {S.}~\bibnamefont
  {Nadj-Perge}}, \bibinfo {author} {\bibfnamefont {V.~S.}\ \bibnamefont
  {Pribiag}}, \bibinfo {author} {\bibfnamefont {J.~W.~G.}\ \bibnamefont
  {van~den Berg}}, \bibinfo {author} {\bibfnamefont {K.}~\bibnamefont {Zuo}},
  \bibinfo {author} {\bibfnamefont {S.~R.}\ \bibnamefont {Plissard}}, \bibinfo
  {author} {\bibfnamefont {E.~P. A.~M.}\ \bibnamefont {Bakkers}}, \bibinfo
  {author} {\bibfnamefont {S.~M.}\ \bibnamefont {Frolov}}, \ and\ \bibinfo
  {author} {\bibfnamefont {L.~P.}\ \bibnamefont {Kouwenhoven}},\ }\href
  {\doibase 10.1103/PhysRevLett.108.166801} {\bibfield  {journal} {\bibinfo
  {journal} {Physical Review Letters}\ }\textbf {\bibinfo {volume} {108}},\
  \bibinfo {pages} {166801} (\bibinfo {year} {2012})}\BibitemShut {NoStop}%
\bibitem [{\citenamefont {Koppens}\ \emph {et~al.}(2007)\citenamefont
  {Koppens}, \citenamefont {Buizert}, \citenamefont {Vink}, \citenamefont
  {Nowack}, \citenamefont {Meunier}, \citenamefont {Kouwenhoven},\ and\
  \citenamefont {Vandersypen}}]{Koppens2007a}%
  \BibitemOpen
  \bibfield  {author} {\bibinfo {author} {\bibfnamefont {F.~H.~L.}\
  \bibnamefont {Koppens}}, \bibinfo {author} {\bibfnamefont {C.}~\bibnamefont
  {Buizert}}, \bibinfo {author} {\bibfnamefont {I.~T.}\ \bibnamefont {Vink}},
  \bibinfo {author} {\bibfnamefont {K.~C.}\ \bibnamefont {Nowack}}, \bibinfo
  {author} {\bibfnamefont {T.}~\bibnamefont {Meunier}}, \bibinfo {author}
  {\bibfnamefont {L.~P.}\ \bibnamefont {Kouwenhoven}}, \ and\ \bibinfo {author}
  {\bibfnamefont {L.~M.~K.}\ \bibnamefont {Vandersypen}},\ }\href {\doibase
  10.1063/1.2722734} {\bibfield  {journal} {\bibinfo  {journal} {Journal of
  Applied Physics}\ }\textbf {\bibinfo {volume} {101}},\ \bibinfo {pages}
  {081706} (\bibinfo {year} {2007})}\BibitemShut {NoStop}%
\bibitem [{\citenamefont {Koppens}\ \emph {et~al.}(2006)\citenamefont
  {Koppens}, \citenamefont {Buizert}, \citenamefont {Tielrooij}, \citenamefont
  {Vink}, \citenamefont {Nowack}, \citenamefont {Meunier}, \citenamefont
  {Kouwenhoven},\ and\ \citenamefont {Vandersypen}}]{Koppens2006S}%
  \BibitemOpen
  \bibfield  {author} {\bibinfo {author} {\bibfnamefont {F.~H.~L.}\
  \bibnamefont {Koppens}}, \bibinfo {author} {\bibfnamefont {C.}~\bibnamefont
  {Buizert}}, \bibinfo {author} {\bibfnamefont {K.~J.}\ \bibnamefont
  {Tielrooij}}, \bibinfo {author} {\bibfnamefont {I.~T.}\ \bibnamefont {Vink}},
  \bibinfo {author} {\bibfnamefont {K.~C.}\ \bibnamefont {Nowack}}, \bibinfo
  {author} {\bibfnamefont {T.}~\bibnamefont {Meunier}}, \bibinfo {author}
  {\bibfnamefont {L.~P.}\ \bibnamefont {Kouwenhoven}}, \ and\ \bibinfo {author}
  {\bibfnamefont {L.~M.~K.}\ \bibnamefont {Vandersypen}},\ }\href {\doibase
  10.1038/nature05065} {\bibfield  {journal} {\bibinfo  {journal} {Nature}\
  }\textbf {\bibinfo {volume} {442}},\ \bibinfo {pages} {766} (\bibinfo {year}
  {2006})}\BibitemShut {NoStop}%
\end{thebibliography}%


\begin{thebibliography}{32}%
\makeatletter
\providecommand \@ifxundefined [1]{%
 \@ifx{#1\undefined}
}%
\providecommand \@ifnum [1]{%
 \ifnum #1\expandafter \@firstoftwo
 \else \expandafter \@secondoftwo
 \fi
}%
\providecommand \@ifx [1]{%
 \ifx #1\expandafter \@firstoftwo
 \else \expandafter \@secondoftwo
 \fi
}%
\providecommand \natexlab [1]{#1}%
\providecommand \enquote  [1]{``#1''}%
\providecommand \bibnamefont  [1]{#1}%
\providecommand \bibfnamefont [1]{#1}%
\providecommand \citenamefont [1]{#1}%
\providecommand \href@noop [0]{\@secondoftwo}%
\providecommand \href [0]{\begingroup \@sanitize@url \@href}%
\providecommand \@href[1]{\@@startlink{#1}\@@href}%
\providecommand \@@href[1]{\endgroup#1\@@endlink}%
\providecommand \@sanitize@url [0]{\catcode `\\12\catcode `\$12\catcode
  `\&12\catcode `\#12\catcode `\^12\catcode `\_12\catcode `\%12\relax}%
\providecommand \@@startlink[1]{}%
\providecommand \@@endlink[0]{}%
\providecommand \url  [0]{\begingroup\@sanitize@url \@url }%
\providecommand \@url [1]{\endgroup\@href {#1}{\urlprefix }}%
\providecommand \urlprefix  [0]{URL }%
\providecommand \Eprint [0]{\href }%
\providecommand \doibase [0]{http://dx.doi.org/}%
\providecommand \selectlanguage [0]{\@gobble}%
\providecommand \bibinfo  [0]{\@secondoftwo}%
\providecommand \bibfield  [0]{\@secondoftwo}%
\providecommand \translation [1]{[#1]}%
\providecommand \BibitemOpen [0]{}%
\providecommand \bibitemStop [0]{}%
\providecommand \bibitemNoStop [0]{.\EOS\space}%
\providecommand \EOS [0]{\spacefactor3000\relax}%
\providecommand \BibitemShut  [1]{\csname bibitem#1\endcsname}%
\let\auto@bib@innerbib\@empty
\bibitem [{\citenamefont {Loss}\ and\ \citenamefont
  {DiVincenzo}(1998)}]{Loss1998}%
  \BibitemOpen
  \bibfield  {author} {\bibinfo {author} {\bibfnamefont {D.}~\bibnamefont
  {Loss}}\ and\ \bibinfo {author} {\bibfnamefont {D.~P.}\ \bibnamefont
  {DiVincenzo}},\ }\href {\doibase 10.1103/PhysRevA.57.120} {\bibfield
  {journal} {\bibinfo  {journal} {Physical Review A}\ }\textbf {\bibinfo
  {volume} {57}},\ \bibinfo {pages} {120} (\bibinfo {year} {1998})}\BibitemShut
  {NoStop}%
\bibitem [{\citenamefont {Hanson}\ \emph {et~al.}(2007)\citenamefont {Hanson},
  \citenamefont {Petta}, \citenamefont {Tarucha},\ and\ \citenamefont
  {Vandersypen}}]{Hanson2007}%
  \BibitemOpen
  \bibfield  {author} {\bibinfo {author} {\bibfnamefont {R.}~\bibnamefont
  {Hanson}}, \bibinfo {author} {\bibfnamefont {J.~R.}\ \bibnamefont {Petta}},
  \bibinfo {author} {\bibfnamefont {S.}~\bibnamefont {Tarucha}}, \ and\
  \bibinfo {author} {\bibfnamefont {L.~M.~K.}\ \bibnamefont {Vandersypen}},\
  }\href {\doibase 10.1103/RevModPhys.79.1217} {\bibfield  {journal} {\bibinfo
  {journal} {Reviews of Modern Physics}\ }\textbf {\bibinfo {volume} {79}},\
  \bibinfo {pages} {1217} (\bibinfo {year} {2007})}\BibitemShut {NoStop}%
\bibitem [{\citenamefont {Petta}\ \emph {et~al.}(2005)\citenamefont {Petta},
  \citenamefont {Johnson}, \citenamefont {Taylor}, \citenamefont {Laird},
  \citenamefont {Yacoby}, \citenamefont {Lukin}, \citenamefont {Marcus},
  \citenamefont {Hanson},\ and\ \citenamefont {Gossard}}]{Petta2005a}%
  \BibitemOpen
  \bibfield  {author} {\bibinfo {author} {\bibfnamefont {J.~R.}\ \bibnamefont
  {Petta}}, \bibinfo {author} {\bibfnamefont {A.~C.}\ \bibnamefont {Johnson}},
  \bibinfo {author} {\bibfnamefont {J.~M.}\ \bibnamefont {Taylor}}, \bibinfo
  {author} {\bibfnamefont {E.~A.}\ \bibnamefont {Laird}}, \bibinfo {author}
  {\bibfnamefont {A.}~\bibnamefont {Yacoby}}, \bibinfo {author} {\bibfnamefont
  {M.~D.}\ \bibnamefont {Lukin}}, \bibinfo {author} {\bibfnamefont {C.~M.}\
  \bibnamefont {Marcus}}, \bibinfo {author} {\bibfnamefont {M.~P.}\
  \bibnamefont {Hanson}}, \ and\ \bibinfo {author} {\bibfnamefont {A.~C.}\
  \bibnamefont {Gossard}},\ }\href {\doibase 10.1126/science.1116955}
  {\bibfield  {journal} {\bibinfo  {journal} {Science (New York, N.Y.)}\
  }\textbf {\bibinfo {volume} {309}},\ \bibinfo {pages} {2180} (\bibinfo {year}
  {2005})}\BibitemShut {NoStop}%
\bibitem [{\citenamefont {Simmons}\ \emph {et~al.}(2011)\citenamefont
  {Simmons}, \citenamefont {Prance}, \citenamefont {{Van Bael}}, \citenamefont
  {Koh}, \citenamefont {Shi}, \citenamefont {Savage}, \citenamefont {Lagally},
  \citenamefont {Joynt}, \citenamefont {Friesen}, \citenamefont {Coppersmith},\
  and\ \citenamefont {Eriksson}}]{Simmons2011}%
  \BibitemOpen
  \bibfield  {author} {\bibinfo {author} {\bibfnamefont {C.}~\bibnamefont
  {Simmons}}, \bibinfo {author} {\bibfnamefont {J.~R.}\ \bibnamefont {Prance}},
  \bibinfo {author} {\bibfnamefont {B.~J.}\ \bibnamefont {{Van Bael}}},
  \bibinfo {author} {\bibfnamefont {T.~S.}\ \bibnamefont {Koh}}, \bibinfo
  {author} {\bibfnamefont {Z.}~\bibnamefont {Shi}}, \bibinfo {author}
  {\bibfnamefont {D.~E.}\ \bibnamefont {Savage}}, \bibinfo {author}
  {\bibfnamefont {M.~G.}\ \bibnamefont {Lagally}}, \bibinfo {author}
  {\bibfnamefont {R.}~\bibnamefont {Joynt}}, \bibinfo {author} {\bibfnamefont
  {M.}~\bibnamefont {Friesen}}, \bibinfo {author} {\bibfnamefont {S.~N.}\
  \bibnamefont {Coppersmith}}, \ and\ \bibinfo {author} {\bibfnamefont {M.~A.}\
  \bibnamefont {Eriksson}},\ }\href {\doibase 10.1103/PhysRevLett.106.156804}
  {\bibfield  {journal} {\bibinfo  {journal} {Physical Review Letters}\
  }\textbf {\bibinfo {volume} {106}},\ \bibinfo {pages} {156804} (\bibinfo
  {year} {2011})}\BibitemShut {NoStop}%
\bibitem [{\citenamefont {Nadj-Perge}\ \emph
  {et~al.}(2010{\natexlab{a}})\citenamefont {Nadj-Perge}, \citenamefont
  {Frolov}, \citenamefont {van Tilburg}, \citenamefont {Danon}, \citenamefont
  {Nazarov}, \citenamefont {Algra}, \citenamefont {Bakkers},\ and\
  \citenamefont {Kouwenhoven}}]{Nadj-Perge2010a}%
  \BibitemOpen
  \bibfield  {author} {\bibinfo {author} {\bibfnamefont {S.}~\bibnamefont
  {Nadj-Perge}}, \bibinfo {author} {\bibfnamefont {S.~M.}\ \bibnamefont
  {Frolov}}, \bibinfo {author} {\bibfnamefont {J.~W.~W.}\ \bibnamefont {van
  Tilburg}}, \bibinfo {author} {\bibfnamefont {J.}~\bibnamefont {Danon}},
  \bibinfo {author} {\bibfnamefont {Y.~V.}\ \bibnamefont {Nazarov}}, \bibinfo
  {author} {\bibfnamefont {R.~E.}\ \bibnamefont {Algra}}, \bibinfo {author}
  {\bibfnamefont {E.~P. A.~M.}\ \bibnamefont {Bakkers}}, \ and\ \bibinfo
  {author} {\bibfnamefont {L.~P.}\ \bibnamefont {Kouwenhoven}},\ }\href
  {\doibase 10.1103/PhysRevB.81.201305} {\bibfield  {journal} {\bibinfo
  {journal} {Physical Review B}\ }\textbf {\bibinfo {volume} {81}},\ \bibinfo
  {pages} {201305} (\bibinfo {year} {2010}{\natexlab{a}})}\BibitemShut
  {NoStop}%
\bibitem [{\citenamefont {Maune}\ \emph {et~al.}(2012)\citenamefont {Maune},
  \citenamefont {Borselli}, \citenamefont {Huang}, \citenamefont {Ladd},
  \citenamefont {Deelman}, \citenamefont {Holabird}, \citenamefont {Kiselev},
  \citenamefont {Alvarado-Rodriguez}, \citenamefont {Ross}, \citenamefont
  {Schmitz}, \citenamefont {Sokolich}, \citenamefont {Watson}, \citenamefont
  {Gyure},\ and\ \citenamefont {Hunter}}]{Maune2012}%
  \BibitemOpen
  \bibfield  {author} {\bibinfo {author} {\bibfnamefont {B.~M.}\ \bibnamefont
  {Maune}}, \bibinfo {author} {\bibfnamefont {M.~G.}\ \bibnamefont {Borselli}},
  \bibinfo {author} {\bibfnamefont {B.}~\bibnamefont {Huang}}, \bibinfo
  {author} {\bibfnamefont {T.~D.}\ \bibnamefont {Ladd}}, \bibinfo {author}
  {\bibfnamefont {P.~W.}\ \bibnamefont {Deelman}}, \bibinfo {author}
  {\bibfnamefont {K.~S.}\ \bibnamefont {Holabird}}, \bibinfo {author}
  {\bibfnamefont {A.~A.}\ \bibnamefont {Kiselev}}, \bibinfo {author}
  {\bibfnamefont {I.}~\bibnamefont {Alvarado-Rodriguez}}, \bibinfo {author}
  {\bibfnamefont {R.~S.}\ \bibnamefont {Ross}}, \bibinfo {author}
  {\bibfnamefont {A.~E.}\ \bibnamefont {Schmitz}}, \bibinfo {author}
  {\bibfnamefont {M.}~\bibnamefont {Sokolich}}, \bibinfo {author}
  {\bibfnamefont {C.~A.}\ \bibnamefont {Watson}}, \bibinfo {author}
  {\bibfnamefont {M.~F.}\ \bibnamefont {Gyure}}, \ and\ \bibinfo {author}
  {\bibfnamefont {A.~T.}\ \bibnamefont {Hunter}},\ }\href {\doibase
  10.1038/nature10707} {\bibfield  {journal} {\bibinfo  {journal} {Nature}\
  }\textbf {\bibinfo {volume} {481}},\ \bibinfo {pages} {344} (\bibinfo {year}
  {2012})}\BibitemShut {NoStop}%
\bibitem [{\citenamefont {Shulman}\ \emph {et~al.}(2012)\citenamefont
  {Shulman}, \citenamefont {Dial}, \citenamefont {Harvey}, \citenamefont
  {Bluhm}, \citenamefont {Umansky},\ and\ \citenamefont
  {Yacoby}}]{Shulman2012}%
  \BibitemOpen
  \bibfield  {author} {\bibinfo {author} {\bibfnamefont {M.~D.}\ \bibnamefont
  {Shulman}}, \bibinfo {author} {\bibfnamefont {O.~E.}\ \bibnamefont {Dial}},
  \bibinfo {author} {\bibfnamefont {S.~P.}\ \bibnamefont {Harvey}}, \bibinfo
  {author} {\bibfnamefont {H.}~\bibnamefont {Bluhm}}, \bibinfo {author}
  {\bibfnamefont {V.}~\bibnamefont {Umansky}}, \ and\ \bibinfo {author}
  {\bibfnamefont {A.}~\bibnamefont {Yacoby}},\ }\href {\doibase
  10.1126/science.1217692} {\bibfield  {journal} {\bibinfo  {journal}
  {Science}\ }\textbf {\bibinfo {volume} {336}},\ \bibinfo {pages} {202}
  (\bibinfo {year} {2012})}\BibitemShut {NoStop}%
\bibitem [{\citenamefont {Bj\"{o}rk}\ \emph {et~al.}(2005)\citenamefont
  {Bj\"{o}rk}, \citenamefont {Fuhrer}, \citenamefont {Hansen}, \citenamefont
  {Larsson}, \citenamefont {Fr\"{o}berg},\ and\ \citenamefont
  {Samuelson}}]{Bjork2005}%
  \BibitemOpen
  \bibfield  {author} {\bibinfo {author} {\bibfnamefont {M.~T.}\ \bibnamefont
  {Bj\"{o}rk}}, \bibinfo {author} {\bibfnamefont {A.}~\bibnamefont {Fuhrer}},
  \bibinfo {author} {\bibfnamefont {A.~E.}\ \bibnamefont {Hansen}}, \bibinfo
  {author} {\bibfnamefont {M.~W.}\ \bibnamefont {Larsson}}, \bibinfo {author}
  {\bibfnamefont {L.~E.}\ \bibnamefont {Fr\"{o}berg}}, \ and\ \bibinfo {author}
  {\bibfnamefont {L.}~\bibnamefont {Samuelson}},\ }\href {\doibase
  10.1103/PhysRevB.72.201307} {\bibfield  {journal} {\bibinfo  {journal}
  {Physical Review B}\ }\textbf {\bibinfo {volume} {72}},\ \bibinfo {pages}
  {201307} (\bibinfo {year} {2005})}\BibitemShut {NoStop}%
\bibitem [{\citenamefont {Pfund}\ \emph
  {et~al.}(2007{\natexlab{a}})\citenamefont {Pfund}, \citenamefont
  {Shorubalko}, \citenamefont {Ensslin},\ and\ \citenamefont
  {Leturcq}}]{Pfund2007a}%
  \BibitemOpen
  \bibfield  {author} {\bibinfo {author} {\bibfnamefont {A.}~\bibnamefont
  {Pfund}}, \bibinfo {author} {\bibfnamefont {I.}~\bibnamefont {Shorubalko}},
  \bibinfo {author} {\bibfnamefont {K.}~\bibnamefont {Ensslin}}, \ and\
  \bibinfo {author} {\bibfnamefont {R.}~\bibnamefont {Leturcq}},\ }\href
  {\doibase 10.1103/PhysRevB.76.161308} {\bibfield  {journal} {\bibinfo
  {journal} {Physical Review B}\ }\textbf {\bibinfo {volume} {76}},\ \bibinfo
  {pages} {161308} (\bibinfo {year} {2007}{\natexlab{a}})}\BibitemShut
  {NoStop}%
\bibitem [{\citenamefont {Nilsson}\ \emph {et~al.}(2009)\citenamefont
  {Nilsson}, \citenamefont {Caroff}, \citenamefont {Thelander}, \citenamefont
  {Larsson}, \citenamefont {Wagner}, \citenamefont {Wernersson}, \citenamefont
  {Samuelson},\ and\ \citenamefont {Xu}}]{Nilsson2009}%
  \BibitemOpen
  \bibfield  {author} {\bibinfo {author} {\bibfnamefont {H.~A.}\ \bibnamefont
  {Nilsson}}, \bibinfo {author} {\bibfnamefont {P.}~\bibnamefont {Caroff}},
  \bibinfo {author} {\bibfnamefont {C.}~\bibnamefont {Thelander}}, \bibinfo
  {author} {\bibfnamefont {M.}~\bibnamefont {Larsson}}, \bibinfo {author}
  {\bibfnamefont {J.~B.}\ \bibnamefont {Wagner}}, \bibinfo {author}
  {\bibfnamefont {L.-E.}\ \bibnamefont {Wernersson}}, \bibinfo {author}
  {\bibfnamefont {L.}~\bibnamefont {Samuelson}}, \ and\ \bibinfo {author}
  {\bibfnamefont {H.~Q.}\ \bibnamefont {Xu}},\ }\href {\doibase
  10.1021/nl901333a} {\bibfield  {journal} {\bibinfo  {journal} {Nano letters}\
  }\textbf {\bibinfo {volume} {9}},\ \bibinfo {pages} {3151} (\bibinfo {year}
  {2009})}\BibitemShut {NoStop}%
\bibitem [{\citenamefont {Nadj-Perge}\ \emph
  {et~al.}(2010{\natexlab{b}})\citenamefont {Nadj-Perge}, \citenamefont
  {Frolov}, \citenamefont {Bakkers},\ and\ \citenamefont
  {Kouwenhoven}}]{Nadj-Perge2010}%
  \BibitemOpen
  \bibfield  {author} {\bibinfo {author} {\bibfnamefont {S.}~\bibnamefont
  {Nadj-Perge}}, \bibinfo {author} {\bibfnamefont {S.~M.}\ \bibnamefont
  {Frolov}}, \bibinfo {author} {\bibfnamefont {E.~P. A.~M.}\ \bibnamefont
  {Bakkers}}, \ and\ \bibinfo {author} {\bibfnamefont {L.~P.}\ \bibnamefont
  {Kouwenhoven}},\ }\href {\doibase 10.1038/nature09682} {\bibfield  {journal}
  {\bibinfo  {journal} {Nature}\ }\textbf {\bibinfo {volume} {468}},\ \bibinfo
  {pages} {1084} (\bibinfo {year} {2010}{\natexlab{b}})}\BibitemShut {NoStop}%
\bibitem [{\citenamefont {Petersson}\ \emph {et~al.}(2012)\citenamefont
  {Petersson}, \citenamefont {McFaul}, \citenamefont {Schroer}, \citenamefont
  {Jung}, \citenamefont {Taylor}, \citenamefont {Houck},\ and\ \citenamefont
  {Petta}}]{Petersson2012}%
  \BibitemOpen
  \bibfield  {author} {\bibinfo {author} {\bibfnamefont {K.~D.}\ \bibnamefont
  {Petersson}}, \bibinfo {author} {\bibfnamefont {L.~W.}\ \bibnamefont
  {McFaul}}, \bibinfo {author} {\bibfnamefont {M.~D.}\ \bibnamefont {Schroer}},
  \bibinfo {author} {\bibfnamefont {M.}~\bibnamefont {Jung}}, \bibinfo {author}
  {\bibfnamefont {J.~M.}\ \bibnamefont {Taylor}}, \bibinfo {author}
  {\bibfnamefont {A.~A.}\ \bibnamefont {Houck}}, \ and\ \bibinfo {author}
  {\bibfnamefont {J.~R.}\ \bibnamefont {Petta}},\ }\href
  {http://arxiv.org/abs/1205.6767} {\  (\bibinfo {year} {2012})},\ \Eprint
  {http://arxiv.org/abs/1205.6767} {arXiv:1205.6767} \BibitemShut {NoStop}%
\bibitem [{\citenamefont {Pfund}\ \emph
  {et~al.}(2007{\natexlab{b}})\citenamefont {Pfund}, \citenamefont
  {Shorubalko}, \citenamefont {Ensslin},\ and\ \citenamefont
  {Leturcq}}]{Pfund2007}%
  \BibitemOpen
  \bibfield  {author} {\bibinfo {author} {\bibfnamefont {A.}~\bibnamefont
  {Pfund}}, \bibinfo {author} {\bibfnamefont {I.}~\bibnamefont {Shorubalko}},
  \bibinfo {author} {\bibfnamefont {K.}~\bibnamefont {Ensslin}}, \ and\
  \bibinfo {author} {\bibfnamefont {R.}~\bibnamefont {Leturcq}},\ }\href
  {\doibase 10.1103/PhysRevLett.99.036801} {\bibfield  {journal} {\bibinfo
  {journal} {Physical Review Letters}\ }\textbf {\bibinfo {volume} {99}},\
  \bibinfo {pages} {036801} (\bibinfo {year} {2007}{\natexlab{b}})}\BibitemShut
  {NoStop}%
\bibitem [{\citenamefont {Rashba}\ and\ \citenamefont
  {Efros}(2003)}]{Rashba2003a}%
  \BibitemOpen
  \bibfield  {author} {\bibinfo {author} {\bibfnamefont {E.~I.}\ \bibnamefont
  {Rashba}}\ and\ \bibinfo {author} {\bibfnamefont {A.~L.}\ \bibnamefont
  {Efros}},\ }\href {\doibase 10.1103/PhysRevLett.91.126405} {\bibfield
  {journal} {\bibinfo  {journal} {Physical Review Letters}\ }\textbf {\bibinfo
  {volume} {91}},\ \bibinfo {pages} {126405} (\bibinfo {year}
  {2003})}\BibitemShut {NoStop}%
\bibitem [{\citenamefont {Golovach}\ \emph {et~al.}(2006)\citenamefont
  {Golovach}, \citenamefont {Borhani},\ and\ \citenamefont
  {Loss}}]{Golovach2006}%
  \BibitemOpen
  \bibfield  {author} {\bibinfo {author} {\bibfnamefont {V.~N.}\ \bibnamefont
  {Golovach}}, \bibinfo {author} {\bibfnamefont {M.}~\bibnamefont {Borhani}}, \
  and\ \bibinfo {author} {\bibfnamefont {D.}~\bibnamefont {Loss}},\ }\href
  {\doibase 10.1103/PhysRevB.74.165319} {\bibfield  {journal} {\bibinfo
  {journal} {Physical Review B}\ }\textbf {\bibinfo {volume} {74}},\ \bibinfo
  {pages} {165319} (\bibinfo {year} {2006})}\BibitemShut {NoStop}%
\bibitem [{\citenamefont {Laird}\ \emph {et~al.}(2007)\citenamefont {Laird},
  \citenamefont {Barthel}, \citenamefont {Rashba}, \citenamefont {Marcus},
  \citenamefont {Hanson},\ and\ \citenamefont {Gossard}}]{Laird2007}%
  \BibitemOpen
  \bibfield  {author} {\bibinfo {author} {\bibfnamefont {E.~A.}\ \bibnamefont
  {Laird}}, \bibinfo {author} {\bibfnamefont {C.}~\bibnamefont {Barthel}},
  \bibinfo {author} {\bibfnamefont {E.~I.}\ \bibnamefont {Rashba}}, \bibinfo
  {author} {\bibfnamefont {C.~M.}\ \bibnamefont {Marcus}}, \bibinfo {author}
  {\bibfnamefont {M.~P.}\ \bibnamefont {Hanson}}, \ and\ \bibinfo {author}
  {\bibfnamefont {A.~C.}\ \bibnamefont {Gossard}},\ }\href {\doibase
  10.1103/PhysRevLett.99.246601} {\bibfield  {journal} {\bibinfo  {journal}
  {Physical Review Letters}\ }\textbf {\bibinfo {volume} {99}},\ \bibinfo
  {pages} {246601} (\bibinfo {year} {2007})}\BibitemShut {NoStop}%
\bibitem [{\citenamefont {Pioro-Ladri\`{e}re}\ \emph
  {et~al.}(2008)\citenamefont {Pioro-Ladri\`{e}re}, \citenamefont {Obata},
  \citenamefont {Tokura}, \citenamefont {Shin}, \citenamefont {Kubo},
  \citenamefont {Yoshida}, \citenamefont {Taniyama},\ and\ \citenamefont
  {Tarucha}}]{Pioro-Ladriere2008}%
  \BibitemOpen
  \bibfield  {author} {\bibinfo {author} {\bibfnamefont {M.}~\bibnamefont
  {Pioro-Ladri\`{e}re}}, \bibinfo {author} {\bibfnamefont {T.}~\bibnamefont
  {Obata}}, \bibinfo {author} {\bibfnamefont {Y.}~\bibnamefont {Tokura}},
  \bibinfo {author} {\bibfnamefont {Y.-S.}\ \bibnamefont {Shin}}, \bibinfo
  {author} {\bibfnamefont {T.}~\bibnamefont {Kubo}}, \bibinfo {author}
  {\bibfnamefont {K.}~\bibnamefont {Yoshida}}, \bibinfo {author} {\bibfnamefont
  {T.}~\bibnamefont {Taniyama}}, \ and\ \bibinfo {author} {\bibfnamefont
  {S.}~\bibnamefont {Tarucha}},\ }\href {\doibase 10.1038/nphys1053} {\bibfield
   {journal} {\bibinfo  {journal} {Nature Physics}\ }\textbf {\bibinfo {volume}
  {4}},\ \bibinfo {pages} {776} (\bibinfo {year} {2008})}\BibitemShut {NoStop}%
\bibitem [{Note1()}]{Note1}%
  \BibitemOpen
  \bibinfo {note} {Due to a small offset in the magnet power supply output, the
  magnetic field scale has been offset to have 0 coincide with the hyperfine
  peak.}\BibitemShut {Stop}%
\bibitem [{\citenamefont {Koppens}\ \emph {et~al.}(2005)\citenamefont
  {Koppens}, \citenamefont {Folk}, \citenamefont {Elzerman}, \citenamefont
  {Hanson}, \citenamefont {van Beveren}, \citenamefont {Vink}, \citenamefont
  {Tranitz}, \citenamefont {Wegscheider}, \citenamefont {Kouwenhoven},\ and\
  \citenamefont {Vandersypen}}]{Koppens2005a}%
  \BibitemOpen
  \bibfield  {author} {\bibinfo {author} {\bibfnamefont {F.~H.~L.}\
  \bibnamefont {Koppens}}, \bibinfo {author} {\bibfnamefont {J.~A.}\
  \bibnamefont {Folk}}, \bibinfo {author} {\bibfnamefont {J.~M.}\ \bibnamefont
  {Elzerman}}, \bibinfo {author} {\bibfnamefont {R.}~\bibnamefont {Hanson}},
  \bibinfo {author} {\bibfnamefont {L.~H.~W.}\ \bibnamefont {van Beveren}},
  \bibinfo {author} {\bibfnamefont {I.~T.}\ \bibnamefont {Vink}}, \bibinfo
  {author} {\bibfnamefont {H.~P.}\ \bibnamefont {Tranitz}}, \bibinfo {author}
  {\bibfnamefont {W.}~\bibnamefont {Wegscheider}}, \bibinfo {author}
  {\bibfnamefont {L.~P.}\ \bibnamefont {Kouwenhoven}}, \ and\ \bibinfo {author}
  {\bibfnamefont {L.~M.~K.}\ \bibnamefont {Vandersypen}},\ }\href {\doibase
  10.1126/science.1113719} {\bibfield  {journal} {\bibinfo  {journal} {Science
  (New York, N.Y.)}\ }\textbf {\bibinfo {volume} {309}},\ \bibinfo {pages}
  {1346} (\bibinfo {year} {2005})}\BibitemShut {NoStop}%
\bibitem [{\citenamefont {Nadj-Perge}\ \emph {et~al.}(2012)\citenamefont
  {Nadj-Perge}, \citenamefont {Pribiag}, \citenamefont {van~den Berg},
  \citenamefont {Zuo}, \citenamefont {Plissard}, \citenamefont {Bakkers},
  \citenamefont {Frolov},\ and\ \citenamefont {Kouwenhoven}}]{Nadj-Perge2012}%
  \BibitemOpen
  \bibfield  {author} {\bibinfo {author} {\bibfnamefont {S.}~\bibnamefont
  {Nadj-Perge}}, \bibinfo {author} {\bibfnamefont {V.~S.}\ \bibnamefont
  {Pribiag}}, \bibinfo {author} {\bibfnamefont {J.~W.~G.}\ \bibnamefont
  {van~den Berg}}, \bibinfo {author} {\bibfnamefont {K.}~\bibnamefont {Zuo}},
  \bibinfo {author} {\bibfnamefont {S.~R.}\ \bibnamefont {Plissard}}, \bibinfo
  {author} {\bibfnamefont {E.~P. A.~M.}\ \bibnamefont {Bakkers}}, \bibinfo
  {author} {\bibfnamefont {S.~M.}\ \bibnamefont {Frolov}}, \ and\ \bibinfo
  {author} {\bibfnamefont {L.~P.}\ \bibnamefont {Kouwenhoven}},\ }\href
  {\doibase 10.1103/PhysRevLett.108.166801} {\bibfield  {journal} {\bibinfo
  {journal} {Physical Review Letters}\ }\textbf {\bibinfo {volume} {108}},\
  \bibinfo {pages} {166801} (\bibinfo {year} {2012})}\BibitemShut {NoStop}%
\bibitem [{sup()}]{supmat}%
  \BibitemOpen
  \href@noop {} {}\bibinfo {note} {See Supplemental Material for details on
  double dot occupation, faster Rabi oscillations, nuclear field and fidelity
  estimates and individual addressing of the qubits.}\BibitemShut {Stop}%
\bibitem [{\citenamefont {Koppens}\ \emph {et~al.}(2006)\citenamefont
  {Koppens}, \citenamefont {Buizert}, \citenamefont {Tielrooij}, \citenamefont
  {Vink}, \citenamefont {Nowack}, \citenamefont {Meunier}, \citenamefont
  {Kouwenhoven},\ and\ \citenamefont {Vandersypen}}]{Koppens2006a}%
  \BibitemOpen
  \bibfield  {author} {\bibinfo {author} {\bibfnamefont {F.~H.~L.}\
  \bibnamefont {Koppens}}, \bibinfo {author} {\bibfnamefont {C.}~\bibnamefont
  {Buizert}}, \bibinfo {author} {\bibfnamefont {K.~J.}\ \bibnamefont
  {Tielrooij}}, \bibinfo {author} {\bibfnamefont {I.~T.}\ \bibnamefont {Vink}},
  \bibinfo {author} {\bibfnamefont {K.~C.}\ \bibnamefont {Nowack}}, \bibinfo
  {author} {\bibfnamefont {T.}~\bibnamefont {Meunier}}, \bibinfo {author}
  {\bibfnamefont {L.~P.}\ \bibnamefont {Kouwenhoven}}, \ and\ \bibinfo {author}
  {\bibfnamefont {L.~M.~K.}\ \bibnamefont {Vandersypen}},\ }\href {\doibase
  10.1038/nature05065} {\bibfield  {journal} {\bibinfo  {journal} {Nature}\
  }\textbf {\bibinfo {volume} {442}},\ \bibinfo {pages} {766} (\bibinfo {year}
  {2006})}\BibitemShut {NoStop}%
\bibitem [{\citenamefont {Nowack}\ \emph {et~al.}(2007)\citenamefont {Nowack},
  \citenamefont {Koppens}, \citenamefont {Nazarov},\ and\ \citenamefont
  {Vandersypen}}]{Nowack2007a}%
  \BibitemOpen
  \bibfield  {author} {\bibinfo {author} {\bibfnamefont {K.~C.}\ \bibnamefont
  {Nowack}}, \bibinfo {author} {\bibfnamefont {F.~H.~L.}\ \bibnamefont
  {Koppens}}, \bibinfo {author} {\bibfnamefont {Y.~V.}\ \bibnamefont
  {Nazarov}}, \ and\ \bibinfo {author} {\bibfnamefont {L.~M.~K.}\ \bibnamefont
  {Vandersypen}},\ }\href {\doibase 10.1126/science.1148092} {\bibfield
  {journal} {\bibinfo  {journal} {Science (New York, N.Y.)}\ }\textbf {\bibinfo
  {volume} {318}},\ \bibinfo {pages} {1430} (\bibinfo {year}
  {2007})}\BibitemShut {NoStop}%
\bibitem [{\citenamefont {Herzog}\ and\ \citenamefont
  {Hahn}(1956)}]{Herzog1956}%
  \BibitemOpen
  \bibfield  {author} {\bibinfo {author} {\bibfnamefont {B.}~\bibnamefont
  {Herzog}}\ and\ \bibinfo {author} {\bibfnamefont {E.}~\bibnamefont {Hahn}},\
  }\href {\doibase 10.1103/PhysRev.103.148} {\bibfield  {journal} {\bibinfo
  {journal} {Physical Review}\ }\textbf {\bibinfo {volume} {103}},\ \bibinfo
  {pages} {148} (\bibinfo {year} {1956})}\BibitemShut {NoStop}%
\bibitem [{\citenamefont {Koppens}\ \emph {et~al.}(2008)\citenamefont
  {Koppens}, \citenamefont {Nowack},\ and\ \citenamefont
  {Vandersypen}}]{Koppens2008}%
  \BibitemOpen
  \bibfield  {author} {\bibinfo {author} {\bibfnamefont {F.~H.~L.}\
  \bibnamefont {Koppens}}, \bibinfo {author} {\bibfnamefont {K.~C.}\
  \bibnamefont {Nowack}}, \ and\ \bibinfo {author} {\bibfnamefont {L.~M.~K.}\
  \bibnamefont {Vandersypen}},\ }\href {\doibase
  10.1103/PhysRevLett.100.236802} {\bibfield  {journal} {\bibinfo  {journal}
  {Physical Review Letters}\ }\textbf {\bibinfo {volume} {100}},\ \bibinfo
  {pages} {236802} (\bibinfo {year} {2008})}\BibitemShut {NoStop}%
\bibitem [{\citenamefont {Reilly}\ \emph {et~al.}(2008)\citenamefont {Reilly},
  \citenamefont {Taylor}, \citenamefont {Laird}, \citenamefont {Petta},
  \citenamefont {Marcus}, \citenamefont {Hanson},\ and\ \citenamefont
  {Gossard}}]{Reilly2008a}%
  \BibitemOpen
  \bibfield  {author} {\bibinfo {author} {\bibfnamefont {D.~J.}\ \bibnamefont
  {Reilly}}, \bibinfo {author} {\bibfnamefont {J.~M.}\ \bibnamefont {Taylor}},
  \bibinfo {author} {\bibfnamefont {E.~A.}\ \bibnamefont {Laird}}, \bibinfo
  {author} {\bibfnamefont {J.~R.}\ \bibnamefont {Petta}}, \bibinfo {author}
  {\bibfnamefont {C.~M.}\ \bibnamefont {Marcus}}, \bibinfo {author}
  {\bibfnamefont {M.~P.}\ \bibnamefont {Hanson}}, \ and\ \bibinfo {author}
  {\bibfnamefont {A.~C.}\ \bibnamefont {Gossard}},\ }\href {\doibase
  10.1103/PhysRevLett.101.236803} {\bibfield  {journal} {\bibinfo  {journal}
  {Physical Review Letters}\ }\textbf {\bibinfo {volume} {101}},\ \bibinfo
  {pages} {236803} (\bibinfo {year} {2008})}\BibitemShut {NoStop}%
\bibitem [{\citenamefont {Tokura}\ \emph {et~al.}(2006)\citenamefont {Tokura},
  \citenamefont {van~der Wiel}, \citenamefont {Obata},\ and\ \citenamefont
  {Tarucha}}]{Tokura2006}%
  \BibitemOpen
  \bibfield  {author} {\bibinfo {author} {\bibfnamefont {Y.}~\bibnamefont
  {Tokura}}, \bibinfo {author} {\bibfnamefont {W.~G.}\ \bibnamefont {van~der
  Wiel}}, \bibinfo {author} {\bibfnamefont {T.}~\bibnamefont {Obata}}, \ and\
  \bibinfo {author} {\bibfnamefont {S.}~\bibnamefont {Tarucha}},\ }\href
  {\doibase 10.1103/PhysRevLett.96.047202} {\bibfield  {journal} {\bibinfo
  {journal} {Physical Review Letters}\ }\textbf {\bibinfo {volume} {96}},\
  \bibinfo {pages} {047202} (\bibinfo {year} {2006})}\BibitemShut {NoStop}%
\bibitem [{\citenamefont {Pryor}\ and\ \citenamefont
  {Flatt\'{e}}(2006)}]{Pryor2006}%
  \BibitemOpen
  \bibfield  {author} {\bibinfo {author} {\bibfnamefont {C.~E.}\ \bibnamefont
  {Pryor}}\ and\ \bibinfo {author} {\bibfnamefont {M.~E.}\ \bibnamefont
  {Flatt\'{e}}},\ }\href {\doibase 10.1103/PhysRevLett.96.026804} {\bibfield
  {journal} {\bibinfo  {journal} {Physical Review Letters}\ }\textbf {\bibinfo
  {volume} {96}},\ \bibinfo {pages} {026804} (\bibinfo {year}
  {2006})}\BibitemShut {NoStop}%
\bibitem [{\citenamefont {Schroer}\ \emph {et~al.}(2011)\citenamefont
  {Schroer}, \citenamefont {Petersson}, \citenamefont {Jung},\ and\
  \citenamefont {Petta}}]{Jung2011a}%
  \BibitemOpen
  \bibfield  {author} {\bibinfo {author} {\bibfnamefont {M.~D.}\ \bibnamefont
  {Schroer}}, \bibinfo {author} {\bibfnamefont {K.~D.}\ \bibnamefont
  {Petersson}}, \bibinfo {author} {\bibfnamefont {M.}~\bibnamefont {Jung}}, \
  and\ \bibinfo {author} {\bibfnamefont {J.~R.}\ \bibnamefont {Petta}},\ }\href
  {\doibase 10.1103/PhysRevLett.107.176811} {\bibfield  {journal} {\bibinfo
  {journal} {Physical Review Letters}\ }\textbf {\bibinfo {volume} {107}},\
  \bibinfo {pages} {176811} (\bibinfo {year} {2011})}\BibitemShut {NoStop}%
\bibitem [{Note2()}]{Note2}%
  \BibitemOpen
  \bibinfo {note} {The Rabi for the $g = 36$ dot was achieved at different gate
  voltages than for the other dot, to improve its visibility, though care was
  taken to measure the intended resonance.}\BibitemShut {Stop}%
\bibitem [{\citenamefont {Burkard}\ \emph {et~al.}(1999)\citenamefont
  {Burkard}, \citenamefont {Loss}, \citenamefont {DiVincenzo},\ and\
  \citenamefont {Smolin}}]{Burkard1999}%
  \BibitemOpen
  \bibfield  {author} {\bibinfo {author} {\bibfnamefont {G.}~\bibnamefont
  {Burkard}}, \bibinfo {author} {\bibfnamefont {D.}~\bibnamefont {Loss}},
  \bibinfo {author} {\bibfnamefont {D.~P.}\ \bibnamefont {DiVincenzo}}, \ and\
  \bibinfo {author} {\bibfnamefont {J.~A.}\ \bibnamefont {Smolin}},\ }\href
  {\doibase 10.1103/PhysRevB.60.11404} {\bibfield  {journal} {\bibinfo
  {journal} {Physical Review B}\ }\textbf {\bibinfo {volume} {60}},\ \bibinfo
  {pages} {11404} (\bibinfo {year} {1999})}\BibitemShut {NoStop}%
\bibitem [{\citenamefont {Meunier}\ \emph {et~al.}(2011)\citenamefont
  {Meunier}, \citenamefont {Calado},\ and\ \citenamefont
  {Vandersypen}}]{Meunier2011}%
  \BibitemOpen
  \bibfield  {author} {\bibinfo {author} {\bibfnamefont {T.}~\bibnamefont
  {Meunier}}, \bibinfo {author} {\bibfnamefont {V.~E.}\ \bibnamefont {Calado}},
  \ and\ \bibinfo {author} {\bibfnamefont {L.~M.~K.}\ \bibnamefont
  {Vandersypen}},\ }\href {\doibase 10.1103/PhysRevB.83.121403} {\bibfield
  {journal} {\bibinfo  {journal} {Physical Review B}\ }\textbf {\bibinfo
  {volume} {83}},\ \bibinfo {pages} {121403} (\bibinfo {year}
  {2011})}\BibitemShut {NoStop}%
\end{thebibliography}%


\begin{thebibliography}{0}%
\makeatletter
\providecommand \@ifxundefined [1]{%
 \@ifx{#1\undefined}
}%
\providecommand \@ifnum [1]{%
 \ifnum #1\expandafter \@firstoftwo
 \else \expandafter \@secondoftwo
 \fi
}%
\providecommand \@ifx [1]{%
 \ifx #1\expandafter \@firstoftwo
 \else \expandafter \@secondoftwo
 \fi
}%
\providecommand \natexlab [1]{#1}%
\providecommand \enquote  [1]{``#1''}%
\providecommand \bibnamefont  [1]{#1}%
\providecommand \bibfnamefont [1]{#1}%
\providecommand \citenamefont [1]{#1}%
\providecommand \href@noop [0]{\@secondoftwo}%
\providecommand \href [0]{\begingroup \@sanitize@url \@href}%
\providecommand \@href[1]{\@@startlink{#1}\@@href}%
\providecommand \@@href[1]{\endgroup#1\@@endlink}%
\providecommand \@sanitize@url [0]{\catcode `\\12\catcode `\$12\catcode
  `\&12\catcode `\#12\catcode `\^12\catcode `\_12\catcode `\%12\relax}%
\providecommand \@@startlink[1]{}%
\providecommand \@@endlink[0]{}%
\providecommand \url  [0]{\begingroup\@sanitize@url \@url }%
\providecommand \@url [1]{\endgroup\@href {#1}{\urlprefix }}%
\providecommand \urlprefix  [0]{URL }%
\providecommand \Eprint [0]{\href }%
\providecommand \doibase [0]{http://dx.doi.org/}%
\providecommand \selectlanguage [0]{\@gobble}%
\providecommand \bibinfo  [0]{\@secondoftwo}%
\providecommand \bibfield  [0]{\@secondoftwo}%
\providecommand \translation [1]{[#1]}%
\providecommand \BibitemOpen [0]{}%
\providecommand \bibitemStop [0]{}%
\providecommand \bibitemNoStop [0]{.\EOS\space}%
\providecommand \EOS [0]{\spacefactor3000\relax}%
\providecommand \BibitemShut  [1]{\csname bibitem#1\endcsname}%
\let\auto@bib@innerbib\@empty
\end{thebibliography}%

\onecolumngrid
\setcounter{figure}{0}
\makeatletter 
\renewcommand{\thefigure}{S\@arabic\c@figure}
\makeatother

\newpage
\section{Supplemental Material to `Fast spin-orbit qubit in an indium antimonide nanowire'}

\section{Double dot occupation}
\begin{figure}[h]
\includegraphics{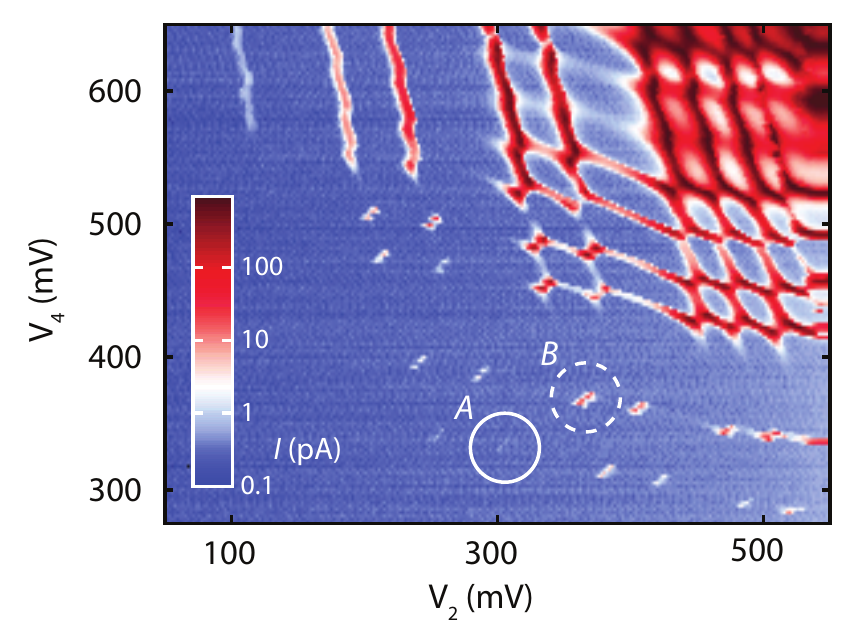}%
\caption{\label{stabdiag}
Stability diagram of our double quantum dot at small positive bias. Circles indicate charge transitions at which data presented in the main paper has been obtained. Transition A (solid circle) indicates the ($2m+1$,$2n+1$)$\rightarrow$($2m$,$2n+2$) transition used for figures 1--3 and transition B (dashed circle) is ($2m+1$,$2n+3$)$\rightarrow$($2m+2$,$2n+2$) used for figure 4).
}
\end{figure}

Figure \ref{stabdiag} displays a stability diagram mapping out the current through the quantum dot as a function of the voltages V$_2$ and V$_4$ on the two plunger gates. The large addition energies and their odd-even structure  indicates that we have reached the few-electron regime. The last visible bias triangles exhibited spin-blockade in the expected positions, further confirming that we have reached the few-electron regime. However, due to the absence of a charge sensor in our device, it is impossible to determine the absolute number of electrons in our dots. Thus, there may be more electrons in the dot for which the charge transitions are not visible (the increased current in the top left of the diagram indicates this is actually likely for one of the dots). We assume, however, that these electrons are paired and can further be ignored. In the labeling of the charge transitions we account for this uncertainty in the electron number by including terms of $2m$ and $2n$ respectively, where $m$ and $n$ are small integers. The first visible triangle is then the ($2m$,$2n+1$)$\rightarrow$($2m+1$,$2n$) transition.

\section{Rabi oscillations}
\begin{figure}[hbt]
\includegraphics{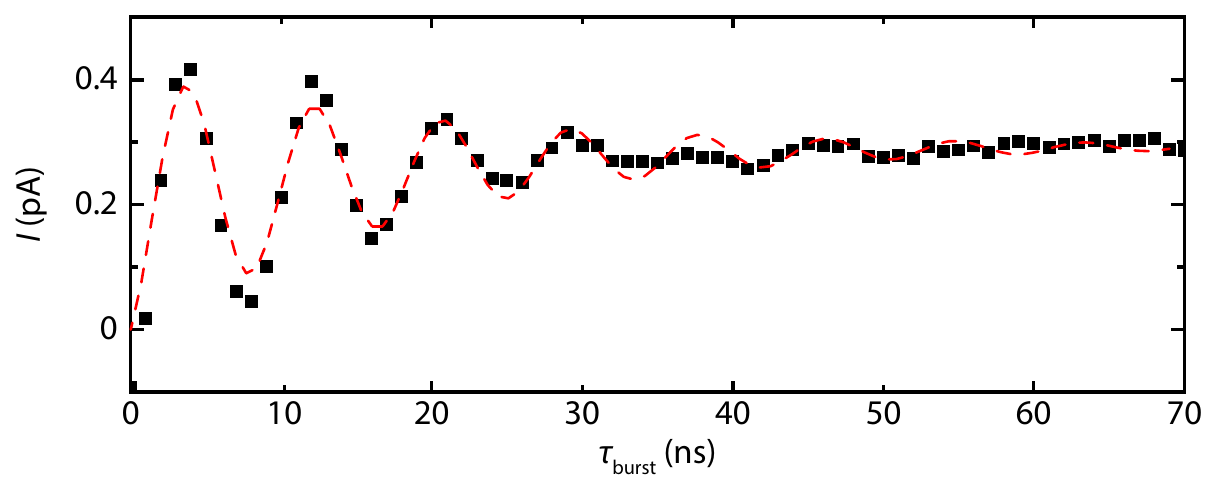}%
\caption{\label{rabi}
Rabi oscillations obtained at the same settings as those presented in the main paper, except for a larger microwave source power of 18~dBm. Dashed line is a fit to a simple exponentially decaying sinusoidal oscillation from which a Rabi frequency of $117\pm1$~MHz is extracted.
}
\end{figure}

In the main text Rabi oscillations have been presented at frequencies up to 104~MHz. Higher Rabi frequencies could be achieved at higher microwave power, at the expense of reduced (i.e. faster decaying) visibility of the oscillations, likely due to photon assisted tunneling. In figure \ref{rabi} we present data displaying this behavior. The higher Rabi frequency of $117\pm1$~MHz only shows half the number of oscillations compared to the data presented in the main text.

\section{Manipulation fidelity}

The nuclear magnetic field changes the intended rotation axis for qubit operations, thereby affecting the fidelity of these rotations. To estimate this fidelity requires knowledge of the spin-orbit driving field $B_{SO}$, as well as the strength of the fluctuating nuclear field $B_N$. The effective field driving the rotations can be determined from the Rabi frequency through $B_{SO} = 2 h f_R / (g \mu_B)$. We estimate the RMS value of the nuclear field fluctuations, $B_N$, from the width of the EDSR peak. More commonly, this is done with the hyperfine peak at $B = 0$ \citeS{Jouravlev2006, Pfund2007S, Danon2009a}. However, it has been previously observed that hyperfine peak widths can vary greatly between different InSb nanowire devices \citeS{Nadj-Perge2012S}, despite similar dot sizes. We therefore make an upper estimate for $B_N$ based on the width of the EDSR peak, which is also broadened by the nuclear field fluctuations at low driving power \citeS{Koppens2007a}.

\begin{figure}
\includegraphics{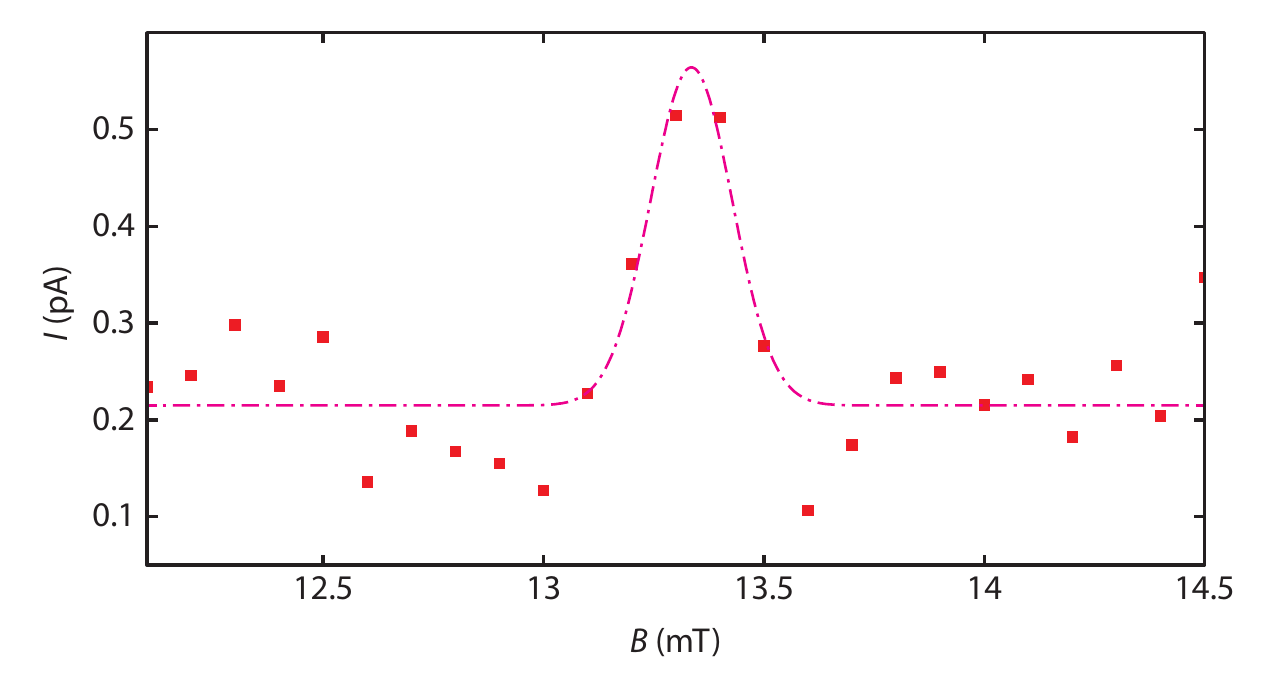}%
\caption{\label{nucfieldfit}
Solid points: data from an EDSR peak obtained at 7.9~GHz driving frequency. Dashed line is a Gaussian fit to the data.
}
\end{figure}

Figure \ref{nucfieldfit} shows an EDSR peak, which has been fitted to a Gaussian curve. The standard deviation of this curve gives the strength of the nuclear field fluctuations in the $z$-direction. From it we determine the total nuclear field fluctuations $B_N = \sqrt{3} B_{N,z} = 0.16 \pm 0.02$~mT. 
Following the procedure explained in \citeS{Koppens2006S} this leads to a manipulation fidelity of $81 \pm 6$~\% for the Rabi oscillations at 104~MHz. 

We note that the absence of a double frequency component in the Rabi oscillations indicates that only one of the qubits is rotated. This cannot be caused by different nuclear fields in the two dots, as $B_N$ is relatively small. This suggests that the second qubit is undriven because of a differing $g$-factor from the first dot and/or a decreased coupling of the microwave driving field to this dot.

\section{Individual addressing of the qubits}

\begin{figure}[hbt]
\includegraphics{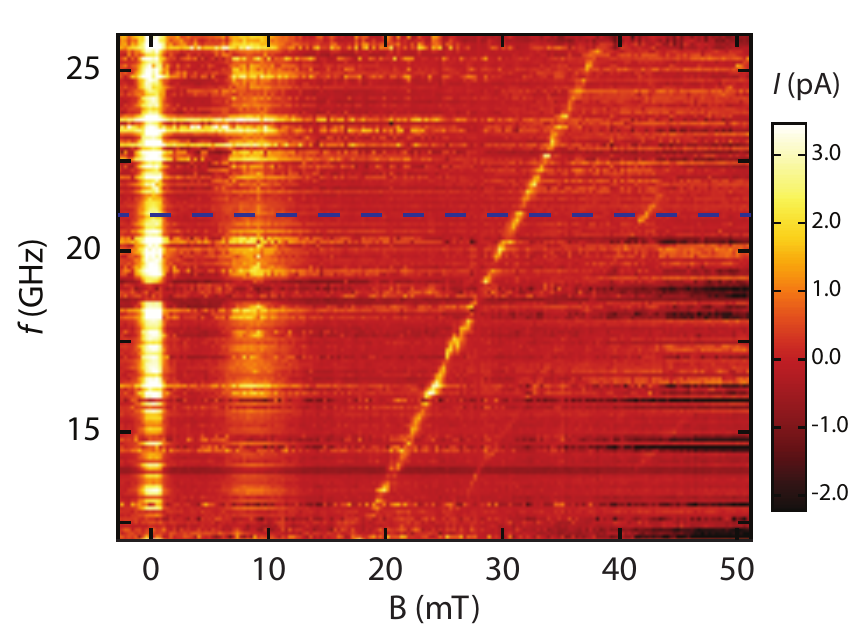}%
\caption{\label{splitedsr}
Magnetic field dependence of the EDSR signals from the two dots. The dashed blue line indicates the position of the trace presented in the main paper. (A vertical linecut has been subtracted to suppress resonances at constant frequency.)
}
\end{figure}

In figure \ref{splitedsr} we present the magnetic field dependence data of the two EDSR peaks already shown in the main paper. The resonance frequency varies linearly with field for both resonances, from which the $g$-factors of 48 and 36 can be determined. A very faint feature may be discerned in the lower right corner of this graph. As it has half the slope of the strongest ($g = 48$) resonance, we attribute it to a multi-photon process.  The bright band that is visible parallel to the hyperfine peak is not caused by the microwave signal, as it is also present without applying microwaves, though it is unclear what its exact origin is.

\bibliographystyleS{apsrev4-1}


%

%

\end{document}